\documentstyle[12pt]{article}
\newcommand{\journal}[4]{{\em #1~}#2\,(19#3)\,#4;}

\newcommand{\pr}{\journal {Phys. Rev.}}

\newcommand{\prl}{\journal {Phys. Rev. Lett.}}

\newcommand{\cmp}{\journal {Comm. Math. Phys.}}

\newcommand{\np}{\journal {Nucl. Phys.}}
\newcommand{\pl}{\journal {Phys. Lett.}}

\newcommand{\prep}{\journal {Phys. Reports}}

\newcommand{\annp}{\journal {Ann. Phys. (N.Y.)}}

\makeatletter
\@addtoreset{equation}{section}
\makeatother

\def\Lp{\displaystyle{\biggl(}}
\def\Rp{\displaystyle{\biggr)}}
\def\LP{\displaystyle{\Biggl(}}
\def\RP{\displaystyle{\Biggr)}}

\newcommand{\complex}{{\kern .1em {\raise .47ex
\hbox {$\scriptscriptstyle |$}}
    \kern -.4em {\rm C}}}
\newcommand{\real}{{{\rm I} \kern -.19em {\rm R}}}
\newcommand{\rational}{{\kern .1em {\raise .47ex
\hbox{$\scripscriptstyle |$}}
    \kern -.35em {\rm Q}}}
\renewcommand{\natural}{{\vrule height 1.6ex width
.05em depth 0ex \kern -.35em {\rm N}}}

\newcommand{\pad}[2]{{\frac{\partial #1}{\partial #2}}}
\newcommand{\fud}[2]  {{\displaystyle{\frac{\delta #1}{\delta #2}}}}

\newcommand{\sla}{\raise.15ex\hbox{$/$}\kern -.57em}

\newcommand{\twiddle}{\lower.9ex\rlap{$\kern -.1em\scriptstyle\sim$}}

\newcommand{\equ}[1]{(\ref{#1})}

\newcommand{\eq}{\begin{equation}}
\newcommand{\eqn}[1]{\label{#1}\end{equation}}
\newcommand{\eea}{\end{eqnarray}}
\newcommand{\eqa}{\begin{eqnarray}}
\newcommand{\eqan}[1]{\label{#1}\end{eqnarray}}
\newcommand{\ba}{\begin{array}}
\newcommand{\ea}{\end{array}}
\newcommand{\eqac}{\begin{equation}\begin{array}{rcl}}
\newcommand{\eqacn}[1]{\end{array}\label{#1}\end{equation}}

\renewcommand{\pad}[2]{{\displaystyle{\frac{\partial #1}{\partial #2}}}}

\newcommand{\intx}{\int d^2 \! x \, }

\begin{document}
\def\ftoday{{\sl  \number\day \space\ifcase\month
\or Janvier\or F\'evrier\or Mars\or avril\or Mai
\or Juin\or Juillet\or Ao\^ut\or Septembre\or Octobre
\or Novembre \or D\'ecembre\fi
\space  \number\year}}


{\large 

\titlepage

\begin{center}

{\huge Infrared and Ultraviolet Finiteness
of Topological BF Theory in Two Dimensions}

\vspace{1cm}

{\Large A. Blasi}\footnote{On leave of absence from Dipartimento di Fisica -
Universit\'a di Trento,
38050 Povo (Trento) - (Italy)}

\vspace{.3cm}
{\it Laboratoire d'Annecy-le-Vieux de Physique de Particules\\
Chemin de Bellevue BP 110, F - 74941 Annecy-le-Vieux Cedex, France}

\vspace{.3cm}
and

\vspace{.3cm}

{\Large N. Maggiore}

\vspace{.3cm}

{\it
Dipartimento di Fisica -- Universit\`a di Genova\\
Istituto Nazionale di Fisica Nucleare -- sez. di Genova\\
Via Dodecaneso, 33 -- 16146 Genova (Italy)}

\end{center}

\vspace{3cm}

\begin{center}
\bf ABSTRACT
\end{center}

{\it The two--dimensional topological BF model is considered in the Landau
gauge in the framework of perturbation theory. Due to the singular behaviour
of the ghost propagator at long distances, a mass term to the ghost fields
is introduced as infrared regulator. Relying on the supersymmetric
algebraic structure of the resulting massive theory, we study the infrared
and ultraviolet renormalizability of the model, with the outcome that it
is perturbatively finite.}

\vfill
GEF-Th-11/1992 \hfill July 1992
\newpage

\section{Introduction}

The BF--systems~\cite{rep} are the only known examples of topological
quantum field theories of the Schwarz type, together with the
three--dimensional
Chern--Simons model, of which they represent the natural extension in an
arbitrary number of spacetime dimensions. From this point of view they allow
to define topological invariants which are the $n$--dimensional generalization
of the $3d$--linking number~\cite{pubper}.

The topological nature of these models has led to the conjecture that, at
least perturbatively, they should be finite. The proof of this assertion
is based on the fact that, besides the gauge invariance, there are additional
symmetries whose algebraic structure becomes transparent in the Landau gauge.
Exploiting this algebra of the supersymmetric type, the conjecture has been
proved for the BF models in three~\cite{ms1} and four~\cite{gms,ms2} flat
euclidean spacetime. Recently, this result has been extended to
di\-men\-sions~\mbox{$n>4$}~\cite{lps}.

What is still lacking is an analogous discussion
concerning the two--dimensional
case, which is particularly interesting because it collects properties of
quantum field theories at first sight very different one from the other.

For instance, it has been noticed in~\cite{rep,blt,wit,sod} that the zero
coupling constant limit of the Yang--Mills action in two dimensions can be
reduced to a topological theory in which the field strenght is coupled to
a scalar field.

The same model is also obtained when considering the Chern--Simons three--form
on a manifold $M_3=S^1\times M_2$, where $S^1$ is a circle and $M_2$ is an
oriented boundaryless surface~\cite{blt,wit}. Indeed, the dimensional reduction
on $M_2$ leads to a metric independent action which can be recognized as a
two--dimensional BF model.

Moreover, this model has been proposed in~\cite{rep,blt,fk} as a gauge
theory of two--dimensional topological quantum gravity.
Indeed, in the particular
case in which the gauge group is $SO(2,1)$, the field equations of motion
describe a torsion--free spin connection living on a manifold with constant
negative curvature. Another interesting feature is the deep relation existing
between the $SO(2,1)$ BF field equations and the dimensionally reduced
solutions of the Hitchin's self--duality equations on a Riemann
surface~\cite{hit}, which provide a mechanism to construct metrics with
constant
negative curvature.

Finally, the equations of motion of the connection for a generic simple
nonabelian gauge group necessarily implies that the scalar field obeys the
constraint typical of a nonlinear sigma model.

Thus, the theory described by a BF model in two dimensions seems to play the
role of a {\it trait d'union} between several field theories, and the question
arises spontaneously wether such a model mantains the finiteness to all
order of perturbation theory.

Although a formal proof of this assertion may be also based on the algebraic
constraints of the model in the Landau gauge, the discussion becomes more
complex due to the singular infrared behaviour of the massless ghost
propagator.
On the other hand, the introduction of a mass term for the ghost fields
breaks the symmetries which are at the root of the procedure to analyze
the perturbative finiteness of the theory.

Therefore the first step consists in giving a mass to the Faddeev--Popov ghosts
by means of a spontaneous symmetry breaking mechanism induced by a suitable
set of external fields.

In this version of the BF system in two dimensions we have massless and massive
propagators and in the discussion of the finiteness of the model we have to
take into account also the infrared canonical dimensions of the
fields~\cite{pr}.

The paper is organized as follows. In section two we describe the classical
model and its supersymmetric algebraic structure. In section three we cure
the infrared singularities of the theory by introducing a mass term for the
ghost fields and we extend the algebraic structure by means of suitable
external
fields. In section four, we show that the symmetries are anomaly--free and that
no counterterms are needed to compensate the divergences of the theory,
{\it i.e.} we prove that the model is perturbatively finite. Finally,
section five is devoted to a discussion of the obtained results.

\section{The massless model}

We consider the metric independent action
\eq
S_{inv} =  {1 \over 2} \int_M d^2 \! x \
   {\ } \varepsilon^{\mu\nu}F^a_{\mu\nu}\phi^a\ ,
\eqn{1.1}
where $M$ is the Euclidean spacetime, $\varepsilon^{\mu\nu}$ is the
completely antisymmetric Levi--Civita tensor $(\varepsilon^{12}=+1)$,
$\phi^a$ is a zero--form, $F^a_{\mu\nu}$ is the usual field strenght
\eq
 F^a_{\mu\nu}=\partial_\mu A^a_\nu - \partial_\nu A^a_\mu +
  f^{abc}A^b_\mu A^c_\nu\
\eqn{1.2}
and $f^{abc}$ are the completely antisymmetric real structure constants of
some nonabelian gauge group $G$, which we assume to be compact and simple.

The action $S_{inv}$ is invariant under the infinitesimal gauge transformations
\eq
\begin{array}{lcl}
\delta A^a_\mu &=& -\left(\partial_\mu\theta^a+f^{abc}A^b_\mu\theta^c
                         \right)\equiv-(D_\mu\theta)^a\\
\delta\phi^a   &=& -f^{abc}\phi^b\theta^c\ ,
\end{array}
\eqn{1.3}
where $\theta^a$ is a local parameter.

The field equations, derived from the action~\equ{1.1}, are
\eq
\fud{S_{inv}}{A^a_\mu} = \varepsilon^{\mu\nu} (D_\nu\phi)^a =0
\eqn{1.4}
\eq
\fud{S_{inv}}{\phi^a} = \frac{1}{2}\varepsilon^{\mu\nu} F^a_{\mu\nu} =0\ .
\eqn{1.5}

The first of the above equations implies that the scalar field $\phi^a$,
in close similarity with a nonlinear $\sigma$--model, lives on a hypersphere
\eq
\phi^a\phi^a=\mbox{constant}\ ,
\eqn{1.6}
while \equ{1.5} yields the vanishing curvature condition typical of topological
models.

Following the BRS procedure to quantize the gauge theories, we introduce a
ghost, an antighost and a Lagrange multiplier field $(c^a,\bar{c}^a,b^a)$ and
we write the nilpotent BRS transformations
\eq\ba{l}
s A^{a}_{\mu}{\ } = - {(D_{\mu}c)}^{a}   \\
s \phi^a = f^{abc}c^b\phi^c \\
s c^{a}{\ }{\ } = {1 \over 2}f^{abc}c^b c^c  \\
s {\bar c}^{a}{\ }{\ } = b^{a}{\ }  \\
s b^{a}=0  \ .
\ea\eqn{1.7}

We then choose the Landau gauge by adding to the action $S_{inv}$ the
gauge--fixing term
\eq\ba{rl}
S_{gf} &\!\! =  \intx  s\ \bar{c}^a\partial A^a \\
&\!\! = \intx \LP b^a\partial A^a - (\partial^\mu\bar{c}^a)(D_\mu c)^a\RP\ .
\ea\eqn{1.8}

In Table 1 we list the canonical dimensions and Faddeev--Popov charges of the
quantum fields

\begin{center}
\begin{tabular}{|l|r|r|r|r|r|}\hline
&$A$ & $\phi$ & $c$ & $\bar{c}$&$b$
\\ \hline
dim&1&0&0&0&0\\ \hline
$\Phi\Pi$&0&0&1&-1&0\\ \hline
\end{tabular}

\vspace{.2cm}{\footnotesize
{\bf Table 1.}Dimensions and Faddeev--Popov charges of the quantum fields.}
\end{center}

The action
\eq
S=S_{inv}+S_{gf}
\eqn{1.9}
is, by construction, BRS invariant
\eq
s S=0\ .
\eqn{1.10}

As a common feature of the topological models in the Landau
gauge~\cite{ms1,gms,ms2,lps,dgs}, the
gauge--fixed action $S$ is left invariant under a further transformation
of the supersymmetric kind and carrying a vectorial index
\eq\ba{l}
\delta_{\mu} A^{a}_{\nu} {\ } = 0 \\
\delta_\mu \phi^a = -\varepsilon_{\mu\nu}\partial^\nu\bar{c}^a \\
\delta_{\mu} c^{a}{\ }{\ }   = - A_{\mu}^{a}            \\
\delta_{\mu} {\bar c}^{a}{\ }{\ }= 0                    \\
\delta_{\mu} b^{a}{\ }{\ } = \partial_{\mu} {\bar c}^{a}    \ .
\ea\eqn{1.11}
Indeed one can easily verify that
\eq
\delta_\mu S =0\ .
\eqn{1.12}

Moreover, the following algebraic structure holds
\eq\ba{l}
\{\ s,s\ \} = 0 \\
\{\ s,\delta_\mu\ \} = \partial_\mu\ +\ \mbox{equations  of  motion}\\
\{\ \delta_\mu,\delta_\nu\ \} =0\ ,
\ea\eqn{1.13}
which closes on--shell on the translations.

In order to write a Slavnov identity and to extend the formalism off--shell,
we couple external sources to the nonlinear BRS variations in~\equ{1.7}
\eq
S_{ext} =   \intx \Lp
        \Omega^{a\mu}( s A^{a}_{\mu} ) + L^{a}( s c^{a} ) +
\rho^a( s \phi^{a} ) \Rp \ ,
\eqn{1.14}
where $(\Omega^{a\mu}, L^a, \rho^a)$ are external sources whose canonical
dimensions and Faddeev--Popov charges are reported in Table 2
\begin{center}
\begin{tabular}{|l|r|r|r|}\hline
&$\Omega$&$L$&$\rho$\\ \hline
dim&1&2&2\\ \hline
$\Phi\Pi$&$-1$&$-2$&$-1$\\ \hline
\end{tabular}

\vspace{.2cm}{\footnotesize {\bf Table 2.}
Dimensions and Faddeev--Popov
charges of the external fields.}
\end{center}

The classical action
\eq
\Sigma = S_{inv} + S_{gf} + S_{ext}
\eqn{1.15}
satisfies the Slavnov identity
\eq
{\cal S}(\Sigma) = 0                           \ ,
\eqn{1.16}
where
\eq
{\cal S}(\Sigma)  =   \intx \LP
       \fud{\Sigma}{\Omega^{a\mu}} \fud{\Sigma}{A^{a}_{\mu}}
   +
       \fud{\Sigma}{\rho^a} \fud{\Sigma}{\phi^a}
   +
       \fud{\Sigma}{L^{a}} \fud{\Sigma}{c^{a}}
   +
       b^{a} \fud{\Sigma}{{\bar c}^{a}}  \RP
\eqn{1.17}
and the corresponding linearized operator
\eq\ba{rl}
B_{\Sigma} = \intx \LP &\!\!
      \fud{\Sigma}{\Omega^{a\mu}} \fud{\ }{A^{a}_{\mu}}
  +
      \fud{\Sigma}{A^{a}_{\mu}} \fud{\ }{\Omega^{a\mu}}
  +
      \fud{\Sigma}{\rho^a}\fud{\ }{\phi^a}
  +
      \fud{\Sigma}{\phi^a}\fud{\ }{\rho^a} \\
&\!\!
  +
      \fud{\Sigma}{L^{a}} \fud{\ }{c^{a}}
  +
      \fud{\Sigma}{c^{a}} \fud{\ }{L^{a}}
  +
      b^{a} \fud{\ }{{\bar c}^{a} }  \RP
\ea\eqn{1.18}
is nilpotent
\eq
B_{\Sigma}B_{\Sigma}= 0       \ .
\eqn{1.19}

Due to the introduction of the source term \equ{1.14}, the
invariance~\equ{1.12}
writes
\eq
{\cal W}_\mu\Sigma=\Delta_\mu\ ,
\eqn{1.20}
where
\eq
{\cal W}_{\mu}  =  \intx \LP
\varepsilon_{\mu\nu}\rho^a    \fud{\ }{A^{a}_{\nu}}
     -\varepsilon_{\mu\nu}
    (\Omega^{a\nu} + \partial^{\nu}{\bar c}^{a})
    \fud{\ }{\phi^a}
     - A^{a}_{\mu} \fud{\ }{c^{a}}
    + (\partial_{\mu}{\bar c}^{a}) \fud{\ }{b^{a}}
    - L^{a} \fud{\ }{\Omega^{a\mu}} \RP
\eqn{1.21}
and the breaking $\Delta_\mu$
\eq
\Delta_{\mu} = \intx \LP
L^{a}(\partial_{\mu}c^{a})-\rho^a\partial_\mu\phi^a
     - \Omega^{a\nu}(\partial_{\mu}A^{a}_{\nu})
- \varepsilon_{\mu\nu}\rho^a\partial^\nu b^a\RP\ ,
\eqn{1.22}
being linear in the quantum fields, is only
present at classical level.

The model, as it happens for all theories in the Landau gauge, has the
further symmetry associated with the integrated ghost equation of
motion~\cite{bps}
\eq
{\cal G}^{a} \Sigma = \Delta^{a}\ ,
\eqn{1.23}
where
\eq
{\cal G}^{a} = \intx {\ }\left({\ }
    { \delta {\ } \over \delta c^{a} }
   +f^{abc}{\bar c}^{b}{ \delta {\ } \over \delta b^{c} } {\ }\right)
\eqn{1.24}
and $\Delta^a$ is also a classical breaking
\eq
\Delta^{a}= \intx {\ }f^{abc}\left({\ }
\Omega^{b\mu}A^c_\mu - L^bc^c + \rho^b\phi^c {\ }\right) \ .
\eqn{1.25}
Anticommuting the ghost equation~\equ{1.23} with the Slavnov
iden\-ti\-ty~\equ{1.16},
one gets the Ward identity for the rigid gauge invariance of the theory
\eq
{\cal H}^a\Sigma =0\ ,
\eqn{1.26}
where
\eq
{\cal H}^a=\sum_{(all{\ }fields{\ }\psi)}\intx\
f^{abc}\psi^b\frac{\delta}{\delta\psi^c}\ ,
\eqn{1.27}
while the anticommutator between the Slavnov operator~\equ{1.17} and the
susy operator~${\cal W}_\mu$~\equ{1.21} expresses the invariance under
translations
\eq
{\cal P}_\mu \Sigma= \sum_{(all{\ }fields{\ }\psi)}
        \intx {\ }(\partial_\mu\psi)
        {\delta {\Sigma}\over \delta \psi}=0\ .
\eqn{1.28}

The set of constraints on the classical action $\Sigma$ is completed by the
gauge condition
\eq
\fud{\Sigma}{b^a} = \partial A^a
\eqn{1.29}
which, commuted with the Slavnov identity~\equ{1.16} yields the antighost
equation of motion
\eq
\fud{\Sigma}{\bar{c}^a}+\partial^\mu\fud{\Sigma}{\Omega^{a\mu}}=0\ .
\eqn{1.30}

By the introduction of the external fields we have reabsorbed the terms
proportional to the equations of motion in~\equ{1.13}; the resulting nonlinear
off--shell algebra for a generic even charged functional $\gamma$ is
\eq\ba{l}
 B_\gamma {\cal S}(\gamma) =0 \\
 \{{\cal W}_\mu,{\cal W}_\nu\}=0 \\
 {\cal W}_\mu{\cal S}(\gamma) + B_\gamma({\cal W}_\mu\gamma-\Delta_\mu) =
{\cal P}_\mu \gamma  \\
 {\cal G}^a{\cal S}(\gamma)+
 B_\gamma({\cal G}^a\gamma-\Delta^a)={\cal H}^a\gamma \\
 {\cal G}^a({\cal W}_\mu\gamma-\Delta_\mu)+
 {\cal W}_\mu({\cal G}^a\gamma-\Delta^a)= 0 \\
 \{{\cal G}^a,{\cal G}^b\} =0 \\
 {[}{\cal H}^a,{\cal G}^b{]}=-f^{abc}{\cal G}^c\\
 {[}{\cal H}^a,{\cal H}^b{]}=-f^{abc}{\cal H}^c\ .
\ea\eqn{1.32}

To summarize, the model we are considering is characterized by

$i)$  the Slavnov identity \equ{1.16};

$ii)$ the susy identity \equ{1.20};

$iii)$  the ghost equation \equ{1.23};

$iv)$ the gauge--fixing condition \equ{1.29}\ ,

and, as by--products, the rigid gauge invariance~\equ{1.26}, the Ward
identity for the translations~\equ{1.28} and the antighost equation~\equ{1.30}.

Now a quantum extension of the model cannot be directly defined from the
action~\equ{1.15} due to the infrared problems in the ghost sector; we shall
see
how these can be solved by the introduction of a mass for the Faddeev--Popov
ghost fields and how the functional differential operators describing the
symmetries modify still preserving the off--shell algebraic structure
in~\equ{1.32}.

\section{Infrared regularization}

An important point to be noticed concerns the infrared behaviour of the
theory. Indeed, the propagators are
\eq
<A^a_\mu\phi^b>=-i\delta^{ab}\frac{\varepsilon_{\mu\nu}k^\nu}{k^2}
\ \ <A^a_\mu b^b>=i\delta^{ab}\frac{k_\mu}{k^2}
\eqn{2.1}
\eq
<\bar{c}^ac^b>=\delta^{ab}\frac{1}{k^2}\ .
\eqn{2.2}
Dealing with a field theory defined on the two--dimensional flat spacetime, we
see that the ghost propagator~\equ{2.2} is not integrable at small
{\it momenta}.
For a correct treatment of the model, we have to regularize it at long
distances and, to do so, we introduce into the action an infrared regulator in
the form of a ghost mass term, so that the ghost propagator becomes
\eq
<\bar{c}^ac^b>_m=\delta^{ab}\frac{1}{k^2+m^2}\ .
\eqn{2.3}

In presence of both massive and massless propagators, we must distinguish
between ultraviolet ($d$) and infrared ($r$)
dimensions of the fields and composite
operators of the model, which coincide in the completely massless
case~\cite{pr}. From
the massive propagator~\equ{2.3} we obtain
\eq
r(\bar{c}+c)=r(\bar{c})+r(c)=2\ ,
\eqn{2.4}
and we have the freedom to choose
\eq
r(\bar{c})=0\ \ ;\ \ r(c)=2\ .
\eqn{2.5}

We summarize the ultraviolet and infrared dimensions of the quantum fields
and sources of the model in Table 3.
\begin{center}
\begin{tabular}{|l|r|r|r|r|r|r|r|r|}\hline
&$A$ & $\phi$ & $c$ & $\bar{c}$&$b$&$\Omega$&$L$&$\rho$
\\ \hline
$d$&1&0&0&0&0&1&2&2\\ \hline
$r$&1&0&2&0&0&1&2&2\\ \hline
\end{tabular}

\vspace{.2cm}{\footnotesize
{\bf Table 3.}Ultraviolet ($d$) and infrared ($r$) dimensions.}
\end{center}

Of course the presence of a ghost mass term may in principle destroy the
algebraic structure~\equ{1.32}, in particular it breaks both
the Slavnov identity~\equ{1.16} and the susy~\equ{1.20}, but we shall see
that we can recover the same algebra in terms of modified operators obtained at
the price of adding a few external sources.

Indeed, let us consider the following term
\eq\ba{rl}
S_m=\intx\LP&\!\!
(\tau_1+m^2)\bar{c}^ac^a + \tau_2(b^ac^a-\frac{1}{2}f^{abc}\bar{c}^ac^bc^c)\\
&\!\!
+\tau_3^\mu\bar{c}^aA^a_\mu +
\tau_4^\mu\left (b^aA^a_\mu+\bar{c}^a(D_\mu c)^a\right)\RP\ ,
\ea\eqn{2.6}
where $(\tau_1,\tau_2,\tau_3^\mu,\tau_4^\mu)$ are external fields whose
ultraviolet and infrared dimensions, as well as their Faddeev--Popov charges
are displayed in Table~4.
\begin{center}
\begin{tabular}{|l|r|r|r|r|}\hline
&$\tau_1$ & $\tau_2$ & $\tau_3$ & $\tau_4$
\\ \hline
$d$&2&2&1&1\\ \hline
$r$&2&2&2&2\\ \hline
$\Phi\Pi$&0&$-1$&1&0\\ \hline
\end{tabular}

\vspace{.2cm}{\footnotesize
{\bf Table 4.}Quantum numbers of the $\tau$--sources.}
\end{center}

The action
\eq
\Sigma_m=\Sigma+S_m
\eqn{2.7}
satisfies the modified Slavnov identity
\eq
\widetilde{\cal S}(\Sigma_m)=0\ ,
\eqn{2.8}
where
\eq\ba{rl}
\widetilde{\cal S}(\Sigma_m)  =   \intx \LP &\!\!
       \fud{\Sigma_m}{\Omega^{a\mu}} \fud{\Sigma_m}{A^{a}_{\mu}}
   +
       \fud{\Sigma_m}{\rho^a} \fud{\Sigma_m}{\phi^a}
   +
       \fud{\Sigma_m}{L^{a}} \fud{\Sigma_m}{c^{a}}
   +
       b^{a} \fud{\Sigma_m}{{\bar c}^{a}}  \\
&\!\!
   -
       (\tau_1+m^2)\fud{\Sigma_m}{\tau_2}
   +
       \tau_3^\mu\fud{\Sigma_m}{\tau_4^\mu} \RP\ ,
\ea\eqn{2.9}
whose corresponding linearized operator
\eq\ba{rl}
\widetilde{B}_{\Sigma_m} = \intx \LP &\!\!
      \fud{\Sigma_m}{\Omega^{a\mu}} \fud{\ }{A^{a}_{\mu}}
  +
      \fud{\Sigma_m}{A^{a}_{\mu}} \fud{\ }{\Omega^{a\mu}}
  +
      \fud{\Sigma_m}{\rho^a}\fud{\ }{\phi^a}
  +
      \fud{\Sigma_m}{\phi^a}\fud{\ }{\rho^a}
  +
      \fud{\Sigma_m}{L^{a}} \fud{\ }{c^{a}}\\
&\!\!
  +
      \fud{\Sigma_m}{c^{a}} \fud{\ }{L^{a}}
  +
      b^{a} \fud{\ }{{\bar c}^{a} }
  -
      (\tau_1+m^2)\fud{\ }{\tau_2}
  +\tau_3^\mu\fud{\ }{\tau_4^\mu} \RP
\ea\eqn{2.9bis}
is still nilpotent
\eq
\widetilde{B}_{\Sigma_m}\widetilde{B}_{\Sigma_m}= 0       \ .
\eqn{2.10}
Similarly, the susy identity~\equ{1.20} now becomes
\eq
\widetilde{\cal W}_\mu\Sigma_m=\widetilde\Delta_\mu\ ,
\eqn{2.11}
where
\eq\ba{rl}
\widetilde{\cal W}_\mu\Sigma_m=\intx \LP &\!\!
\varepsilon_{\mu\nu}\rho^a    \fud{\ }{A^{a}_{\nu}}
     -\varepsilon_{\mu\nu}
    (\Omega^{a\nu} + \partial^{\nu}{\bar c}^{a} -\tau_4^\nu\bar{c}^a)
    \fud{\ }{\phi^a}
     - A^{a}_{\mu} \fud{\ }{c^{a}}  \\&\!\!
    + (\partial_{\mu}{\bar c}^{a}) \fud{\ }{b^{a}}
    - L^{a} \fud{\ }{\Omega^{a\mu}}
    -(\partial_\mu\tau_2)\fud{\ }{\tau_1}\\&\!\!
    +\left [ \partial_\mu\tau_4^\nu-\delta_\mu^\nu(\tau_1+m^2)\right]
       \fud{\ }{\tau_3^\nu}-\tau_2\fud{\ }{\tau_4^\mu}\RP
\ea\eqn{2.12}
and
\eq\ba{rl}
\widetilde\Delta_{\mu} = \intx \LP &\!\!
L^{a}(\partial_{\mu}c^{a})-\rho^a\partial_\mu\phi^a
     - \Omega^{a\nu}(\partial_{\mu}A^{a}_{\nu})
- \varepsilon_{\mu\nu}\rho^a\partial^\nu b^a\\
&\!\!
+\varepsilon_{\mu\nu}\rho^a\tau_3^\nu\bar{c}^a+
\varepsilon_{\mu\nu}\rho^a\tau_4^\nu b^a\RP\ .
\ea\eqn{2.12bis}

In addition, the dependence of the classical action $\Sigma_m$ on the massive
parameter $m^2$ is controlled by the following shift equation
\eq
\LP\pad{\ }{m^2}-\intx\fud{\ }{\tau_1}\RP\Sigma_m=0\ .
\eqn{shift}

The gauge condition~\equ{1.29} reads now
\eq
\fud{\Sigma_m}{b^a}=\partial A^a+\tau_2 c^a+\tau_4^\mu A^a_\mu\ ,
\eqn{2.13}
which, commuted with the Slavnov identity~\equ{2.8} gives the new
``antighost'' equation of motion
\eq
\bar{\cal G}^a(x)\Sigma_m=\Delta^a_{(g)}(x)\ ,
\eqn{2.14}
where
\eq
\bar{\cal G}^a(x)=\fud{\ }{\bar{c}^a}+\partial^\mu\fud{\ }{\Omega^{a\mu}}
+\tau_4^\mu\fud{\ }{\Omega^{a\mu}}-\tau_2\fud{\ }{L^a}\ ,
\eqn{2.15}
and $\Delta^a_{(g)}(x)$ is the classical breaking given by
\eq
\Delta^a_{(g)}(x)=(\tau_1+m^2)c^a-\tau_3^\mu A^a_\mu\ .
\eqn{2.16}

The ghost equation~\equ{1.23} acquires also an additional linear breaking
\eq
{\cal G}^a\Sigma_m=\widetilde\Delta^a\ ,
\eqn{2.17}
where
\eq
\widetilde\Delta^a=\Delta^a-\intx\Lp
(\tau_1+m^2)\bar{c}^a+\tau_2b^a\Rp\ .
\eqn{2.18}

One can verify that the algebraic structure \equ{1.32}, written for the
modified
operators, survives unaltered.

At this stage we can analyze the perturbative renormalizability of the
massive model whose classical action is $\Sigma_m$ in~\equ{2.7}. The
infared regulator $m^2$ and the satellite external fields needed to implement
the breaking to the original symmetry, affect the ghost sector and we expect
that there are no alterations induced by $m^2$ in the topological properties
of the original massless model.

\section{Perturbative finiteness}

We prove the perturbative finiteness of the model by first showing the absence
of counterterms, leaving as a next step the proof that the symmetries
are not anomalous. Since we have a theory with both massive and
massless propagators, we must take into account either the ultraviolet or the
infrared dimensions of the possible counterterms arising from the analysis
of the stability and of the absence of anomalies. Indeed,  renormalizability
of the theory requires the counterterms to be integrated local functionals
with vanishing ghost number, ultraviolet dimensions $d\leq 2$ and infrared
dimensions $r\geq 2$. In particular, we must check the absence of counterterms
with infrared dimensions $r<2$, which otherwise introduce incurable
long--distance divergences in the model~\cite{pr}.

\subsection{Absence of counterterms}

For what concerns the stability of the model, we shall find that no counterterm
is compatible with the algebraic constraints; indeed a local perturbation
$\Sigma^c$ of the classical action $\Sigma_m$ obeys
\eq
\fud{\Sigma^c}{b^a(x)}=0\ ,
\eqn{3.1}
\eq
{\cal G}^a\Sigma^c =0\ ,
\eqn{3.2}
\eq
\bar{\cal G}^a(x)\Sigma^c=0\ ,
\eqn{3.4}
\eq
\widetilde{\cal W}_\mu\Sigma^c=0\ ,
\eqn{3.5}
\eq
\widetilde{B}_{\Sigma_m}\Sigma^c=0\ ,
\eqn{3.6}
moreover we impose
\eq
\left.\fud{\Sigma^c}{\tau_1(x)}\right |_{\psi=0}=0\ ,\ \ \ \psi =
\mbox{all fields}\ ,
\eqn{3.7}
since the mass to the ghost fields in \equ{2.6} is provided by the spontaneous
symmetry breaking in the direction of the~$\tau_1$--external field.

We first analyze \equ{3.1}--\equ{3.4}; condition \equ{3.1}
is satisfied by a functional which does not depend on $b^a$, while the
ghost condition~\equ{3.2} implies that $\Sigma^c$ does not depend on the
undifferentiated ghost field $c^a$. Finally, the most general functional
having ultraviolet dimensions $d\leq 2$ and vanishing Faddeev--Popov charge
which obeys the stability conditions~\equ{3.1}--\equ{3.4} and~\equ{3.7}
is
\eq
\Sigma^c=\Sigma^c_{(0)}+\Sigma^c_{(2)}\ ,
\eqn{3.9}
where, according to the ultraviolet dimensions,
\eq
\Sigma^c_{(0)}=\sum_{n=2}^{+\infty}
\intx a_0^{P_n}\Phi^{P_n}
\eqn{3.10}
and
\eq\ba{rl}
\Sigma^c_{(2)} = \intx \sum_{n=0}^{+\infty}\LP&\!\!
a_1^{(ab)P_n} (\partial^\mu\phi^a)(\partial_\mu\phi^b) +
a_2^{[ab]P_n} \varepsilon^{\mu\nu}(\partial_\mu\phi^a)(\partial_\nu\phi^b)
\\&\!\!+
a_3^{abP_n} A^a_\mu(\partial^\mu\phi^b) +
a_4^{abP_n} \varepsilon^{\mu\nu}A^a_\mu(\partial_\nu\phi^b) +
a_5^{aP_n} \tau_4^\mu(\partial_\mu\phi^a)
\\&\!\!+
a_6^{aP_n} \varepsilon^{\mu\nu}\tau_{4\mu}(\partial_\nu\phi^a)+
a_7^{(ab)P_n} A^{a\mu}A^a_\mu +
a_8^{[ab]P_n} \varepsilon^{\mu\nu}A^a_\mu A^b_\nu
\\&\!\!+
a_9^{aP_n} \tau_4^\mu A^a_\mu +
a_{10}^{aP_n} \varepsilon^{\mu\nu}\tau_{4\mu}A^a_\nu +
a_{11}^{abP_n} \widehat\Omega^{a\mu}(\partial_\mu c^b)
\\&\!\!+
a_{12}^{abP_n} \varepsilon^{\mu\nu}\widehat\Omega^a_\mu(\partial_\nu c^b) +
a_{13}^{aP_n} \tau_3^\mu \widehat\Omega^a_\mu+
a_{14}^{aP_n} \varepsilon^{\mu\nu}\tau_{3\mu}\widehat\Omega^a_\nu
\\&\!\!+
a_{15}^{P_n} \tau_4^\mu\tau_{4\mu} +
a_{16}^{(aP_n)} \tau_1\phi^a+
a_{17}^{(aP_n)} m^2\phi^a\RP\Phi^{P_n}\ ,
\ea\eqn{3.11}
with the short--hand notation
\eq
P_n\equiv (p_1\ldots p_n)\ \ ;\ \ \Phi^{P_n}\equiv\phi^{p_1}\ldots\phi^{p_n}\ ,
\eqn{3.12}
and $(a_0,\ldots,a_{17})$ are invariant tensors. Furthermore
\eq
a_2^{[ab](p_1\ldots p_n)} + \frac{1}{2}\sum_{k=1}^n\LP
a_2^{[ap_k](bp_1\ldots\hat{p}_k\ldots p_n)} +
a_2^{[p_kb](ap_1\ldots\hat{p}_k\ldots p_n)} \RP=0\ ,
\eqn{3.13}
where the hat means omission of the relative index. Moreover,
condition~\equ{3.4} is solved by a the counterterm~$\Sigma^c$ depending
on the combinations
\eq\ba{rcl}
\widehat\Omega^{a\mu} &=& \Omega^{a\mu} +\partial^\mu\bar{c}^a-
                \tau_4^\mu\bar{c}^a\\
\widehat{L}^a &=& L^a -\tau_2\bar{c}^a\ .
\ea\eqn{3.14}

Now, it is sufficient to analyze the susy condition~\equ{3.6} to conclude
that~$\Sigma^c$ vanishes, {\it i.e.}
\eq
\widetilde{\cal W}_\mu\Sigma^c=0\Longrightarrow\Sigma^c=0\ .
\eqn{3.15}

As it happens for the Chern--Simons theory and for the BF models in
higher dimensions~\cite{ms1,gms,ms2}, there is no need of the Slavnov
condition~\equ{3.6}
to show that these so--called Schwarz--type topological quantum field theories
admit no counterterm. This is mainly due to
the presence of the susy~\equ{2.11} and of the ghost
equation~\equ{2.17}, without which other techniques
must be used~\cite{dlps} to reach the same goal~\equ{3.15}. We do not report
here the algebraic discussion of~\equ{3.15}, which is as straightforward as
it is lenghtly and tedious.

\subsection{Absence of anomalies}

To complete the proof of the perturbative finiteness of the model,
we must show that the symmetries are not anomalous. Their quantum
implementation must be discussed having in mind that we must control also the
infrared dimensionality of the term compensating the breaking; this is due to
the fact that renormalizability by power counting forbids the presence of
counterterms having infared dimensions strictly less than two.

There is no problem for the quantum extensions of the gauge
condition~\equ{2.13} and of
the antighost equation~\equ{2.14}, which, written for the quantum vertex
functional~$\Gamma$, read
\eq
\fud{\Gamma}{b^a(x)}=\Delta^a(x)
\eqn{3.16}
\eq
\bar{\cal G}^a(x)=\bar\Delta^a(x)\ .
\eqn{3.17}
The Quantum Action Principle (Q.A.P.)~\cite{qap} insures that the breakings
$\Delta^a(x)$ and $\bar\Delta^a(x)$ are local functionals with the quantum
numbers in Table~5
\begin{center}
\begin{tabular}{|l|r|r|}\hline
&$\Delta$ & $\bar\Delta$
\\ \hline
$d$&$\leq2$&$\leq2$\\ \hline
$r$&$\geq2$&$\geq2$\\ \hline
$\Phi\Pi$&0&$+1$\\ \hline
\end{tabular}

\vspace{.2cm}{\footnotesize
{\bf Table 5.}Quantum numbers of $\Delta^a$, $\bar\Delta^a$.}
\end{center}

In particular, both $\Delta^a$ and $\bar\Delta^a$ have infrared dimensions
greater than or equal to two, and they can easily be reabsorbed by counterterms
with the correct infrared and ultraviolet dimensions
\eq
\Delta^a(x)=\fud{}{b^a(x)}\Delta\ \ ;\ \
d(\Delta)\leq 2\leq r(\Delta)
\eqn{3.18}
\eq
\bar\Delta^a(x)=\bar{\cal G}^a(x)\bar\Delta\ \ ;\ \
d(\bar\Delta)\leq 2\leq r(\bar\Delta)\ .
\eqn{3.19}

The nature of the problems that might arise when discussing a mixed theory
can be tasted in the quantum extension of the ghost equation~\equ{2.17},
which reads
\eq
{\cal G}^a\Gamma=\Delta^a\ .
\eqn{3.20}
According to the Q.A.P., the breaking $\Delta^a$ is an integrated local
functional with ghost number~$-1$, ultraviolet dimension $d(\Delta^a)\leq 2$
and infrared dimension $r(\Delta^a)\geq 0$. In particular, the Q.A.P. does not
forbid a breaking with vanishing infrared dimension
\eq
\Delta^a_{(0)}=\intx\bar{c}^aP(\phi)\ ,
\eqn{3.21}
where $P(\phi)$ is a polinomial in the scalar field~$\phi^a$. But such a
breaking is not allowed by the antighost equation, under which the breaking
$\Delta^a$ must be invariant. One can easily verify that
\eq
\Delta^a={\cal G}^a\Delta\ \ ;\ \ d(\Delta)\leq 2\leq r(\Delta)\ .
\eqn{3.22}

To summarize, we assume that the quantum vertex functional~$\Gamma$ satisfies
\eqa
\fud{\Gamma}{b^a} &=& \partial A^a +\tau_2c^a+\tau_4^\mu A^a_\mu\nonumber\\
\bar{\cal G}^a(x)\Gamma&=&\Delta^a_{(g)}(x)\label{3.23}\\
{\cal G}^a\Gamma&=&\widetilde\Delta^a\ ,\nonumber
\eea
which respectively are the quantum extension of~\equ{2.13},~\equ{2.14}
and~\equ{2.17}.

It is easy to show that the rigid gauge invariance also holds to all
orders of perturbation theory
\eq
{\cal H}^a\Gamma=0\ .
\eqn{rigid}

To prove that the Slavnov identity~\equ{2.8} and the susy~\equ{2.11}
are not anomalous, we adopt the method illustrated in~\cite{ms2}, which
consists in collecting the symmetries~$s$~\equ{1.7} and~$\delta_\mu$~\equ{1.11}
and the translation operator~${\cal P}_\mu$~\equ{1.28} into a unique
operator~${\cal Q}$
\eq
{\cal Q}\equiv s + \xi^\mu\delta_\mu +\eta^\mu{\cal P}_\mu
-\xi^\mu\pad{}{\eta^\mu}
\eqn{3.25}
by means of two global anticommuting parameters~$\xi^\mu$ and~$\eta^\mu$, whose
ghost numbers are respectively~2 and~1.

The operator {\cal Q} is nilpotent
\eq
{\cal Q}^2=0\ ,
\eqn{3.26}
and describes a symmetry of the gauge--fixed action
\eq
{\cal Q}(S_{inv}+S_{gf})=0\ .
\eqn{3.27}

Once we have modified the source term as follows
\eq\ba{rl}
S^{({\cal Q})}_{ext}=\intx\LP&\!\!
\Omega^{a\mu}\left[-(D_\mu c)^a+\eta^\nu\partial_\nu A^a_\mu\right] +
L^a\left[\frac{1}{2}f^{abc}c^bc^c-\xi^\mu A^a_\mu+\eta^\mu\partial_\mu c^a
\right]\\&\!\!+
\rho^a\left[f^{abc}c^b\phi^c-\xi^\mu\varepsilon_{\mu\nu}
(\partial^\nu\bar{c}^a+\Omega^{a\nu}-\tau_4^\nu\bar{c}^a)
+\eta^\mu\partial_\mu\phi^a\right]\RP\ ,
\ea\eqn{3.28}
the new action
\eq
{\cal I}=S_{inv}+S_{gf}+S^{({\cal Q})}_{ext}+S_m
\eqn{3.29}
satifies the generalized Slavnov identity
\eq
{\cal D}({\cal I})=0\ ,
\eqn{3.30}
where
\eq\ba{rl}
{\cal D}({\cal I})=\intx\LP&\!\!
\fud{{\cal I}}{\Omega^{a\mu}}
\fud{{\cal I}}{A^{a}_{\mu}}
+
\fud{{\cal I}}{\rho^a}
\fud{{\cal I}}{\phi^a}
+
\fud{{\cal I}}{L^a}
\fud{{\cal I}}{c^a}
+
({\cal Q}b^a )
\fud{{\cal I}}{b^a}
\\&\!\!
+
({\cal Q}\bar{c}^a)
\fud{{\cal I}}{{\bar c}^{a}}
+
({\cal Q}\tau_1)
\fud{{\cal I}}{{\tau_1}}
+
({\cal Q}\tau_2)
\fud{{\cal I}}{{\tau_2}}
\\&\!\!
+
({\cal Q}\tau_3^\mu)
\fud{{\cal I}}{{\tau_3^\mu}}
+
({\cal Q}\tau_4^\mu)
\fud{{\cal I}}{{\tau_4^\mu}}  \RP
-\xi^\mu\pad{{\cal I}}{\eta^\mu}\ ,
\ea\eqn{3.31}
whose corresponding linearized Slavnov operator~${\cal D}_{\cal I}$ is
nilpotent
\eq
{\cal D}_{\cal I}{\cal D}_{\cal I}=0\ .
\eqn{3.32}

Besides the generalized Slavnov identity~\equ{3.30}, the action~${\cal I}$
satisfies the symmetries~\equ{3.23},~\equ{rigid} and also
\eq
\pad{\cal I}{\xi^\mu}=\Delta_\mu^{(\xi)}\ \ ;\ \
\pad{\cal I}{\eta^\mu}=\Delta_\mu^{(\eta)}\ ,
\eqn{3.33}
where $\Delta_\mu^{(\xi)}$ and $\Delta_\mu^{(\eta)}$ are linear breakings
\eq
\Delta_\mu^{(\xi)}=-\intx\LP
L^aA^a_\mu+\varepsilon_{\mu\nu}\rho^a\left(
\Omega^{a\nu}+\partial^\nu\bar{c}^a-\tau_4^\nu\bar{c}^a\right)\RP
\eqn{3.34}
\eq
\Delta_\mu^{(\eta)}=-\intx\LP
\Omega^{a\nu}\partial_\mu A^a_\nu-L^a\partial_\mu c^a+\rho^a\partial_\mu\phi^a
\RP\ .
\eqn{3.35}

The following nonlinear algebra
\eq\ba{rcl}
\pad{}{\xi^\mu}{\cal D}(\gamma) &-&
D_\gamma \left(\pad{\gamma}{\xi^\mu}-\Delta^{(\xi)}_\mu\right)=
\left(\widetilde{\cal W}_\mu\gamma-\widetilde\Delta_\mu\right)-
\left(\pad{\gamma}{\eta^\mu}-\Delta^{(\eta)}_\mu\right)\\
&+\intx&\varepsilon_{\mu\nu}\eta^\lambda\LP
(\partial^\nu\rho^a)(\partial_\lambda\bar{c}^a)+
\rho^a\bar{c}^a\partial_\lambda\tau_4^\nu+
\rho^a\tau_4^\nu\partial_\lambda\bar{c}^a\RP\\
&+\intx&\varepsilon_{\mu\nu}\xi^\nu\rho^a\bar{c}^a\tau_2
\ea\eqn{3.36}
\eq
\pad{}{\eta^\mu}{\cal D}(\gamma) +
D_\gamma \left(\pad{\gamma}{\eta^\mu}-\Delta^{(\eta)}_\mu
\right)={\cal P}_\mu\gamma\ ,
\eqn{3.36bis}
holds for any even ghost charged functional $\gamma$. If~$\gamma$ is the
quantum vertex functional~$\Gamma^{({\cal Q})}$  satisfying
\eq
\pad{\Gamma^{({\cal Q})}}{\eta^\mu}=\Delta^{(\eta)}_\mu\ \ ;\ \
\pad{\Gamma^{({\cal Q})}}{\xi^\mu}=\Delta^{(\xi)}_\mu
\eqn{3.37}
\eq
{\cal D}(\Gamma^{({\cal Q})})=0\ ,
\eqn{3.38}
from~\equ{3.36} and \equ{3.36bis} we have
\eq\ba{rl}
\widetilde{\cal W}_\mu\Gamma^{({\cal Q})}=&\!\!
\widetilde\Delta_\mu\\
&\!\!
-\intx\varepsilon_{\mu\nu}
\eta^\lambda\LP
(\partial^\nu\rho^a)(\partial_\lambda\bar{c}^a)+
\rho^a\bar{c}^a\partial_\lambda\tau_4^\nu+
\rho^a\tau_4^\nu\partial_\lambda\bar{c}^a
+\xi^\nu\rho^a\bar{c}^a\tau_2\RP
\ea\eqn{3.39}
\eq
{\cal P}_\mu\Gamma^{({\cal Q})}=0\ .
\eqn{3.40}

In particular, at vanishing global ghosts, the validity of
equations~\equ{3.37} and~\equ{3.38} implies for the quantum vertex functional
\eq
\Gamma\equiv\left.\Gamma^{({\cal Q})}\right|_{\xi=\eta=0}
\eqn{3.41}
the identities
\eq
\widetilde{\cal S}(\Gamma)=0
\eqn{3.42}
\eq
\widetilde{\cal W}_\mu\Gamma=\widetilde\Delta_\mu\ .
\eqn{3.43}

Since the classical constraints~\equ{3.33} are easily implemented at the
quantum level~\equ{3.37}, the extension of the generalized Slavnov
identity~\equ{3.30} to all orders of perturbation theory implies the
identities~\equ{3.42},~\equ{3.43} and the translation quantum invariance,
so that we recover the wanted result.

Let us now prove that the generalized Slavnov identity~\equ{3.30} is
not anomalous.

According to the Q.A.P., the quantum extension of the classical
identity~\equ{3.30} is
\eq
{\cal D}(\Gamma^{({\cal Q})})={\cal A}\cdot\Gamma^{({\cal Q})}=
{\cal A}+O(\hbar{\cal A})\ ,
\eqn{3.44}
where the breaking~${\cal A}\cdot\Gamma^{({\cal Q})}$ is a quantum insertion
whose
lowest nonvanishing order in~$\hbar$,~$\cal A$, is an integrated local
functional with Faddeev--Popov charge~$+1$, ultraviolet dimension
$d({\cal A})\leq 2$ and infrared dimension $r({\cal A})\geq 0$.

The breaking ${\cal A}$ must satisfy the following consistency conditions
\eq
\fud{{\cal A}}{b^a} = 0
\eqn{3.45}
\eq
\bar{\cal G}^a(x){\cal A}=0
\eqn{3.455}
\eq
{\cal G}^a{\cal A}=0
\eqn{3.45bis}
and
\eq
{\cal D}_{\cal I}{\cal A}=0\ .
\eqn{3.46}

The constraints~\equ{3.45}--\equ{3.45bis} imply that
\eq
{\cal A}={\cal A}\left(A^a_\mu, \partial_\mu c^a,\phi^a,
\widehat\Omega^{a\mu},\widehat{L}^a,\rho^a,\tau_1,\tau_2,\tau_3^\mu,
\tau_4^\mu\right)\ ,
\eqn{3.47}
{\it i.e.} the breaking {\cal A} does not depend on~$b^a$ and on the
undifferentiated ghost field~$c^a$ and it depends only on the
combinations~\equ{3.14}. Let us write~${\cal A}$ as
\eq
{\cal A}={\cal A}^{(0)}+{\cal A}^{(1)}+{\cal A}^{(2)}\ ,
\eqn{3.48}
according to the ultraviolet dimensions. Since the linearized Slavnov
operator~${\cal D}_{\cal I}$ does not alter the ultraviolet dimensions,
the consistency condition~\equ{3.46} reads
\eq
{\cal D}_{\cal I}{\cal A}^{(0)}=0
\eqn{3.49}
\eq
{\cal D}_{\cal I}{\cal A}^{(1)}=0
\eqn{3.50}
\eq
{\cal D}_{\cal I}{\cal A}^{(2)}=0\ .
\eqn{3.51}
For what concerns the first two conditions~\equ{3.49} and~\equ{3.50},
we cannot find local functionals of the type~\equ{3.47} having ultraviolet
dimension less than two and Faddeev--Popov charge~$+1$, hence
\eq
{\cal A}^{(0)}={\cal A}^{(1)}=0\ .
\eqn{3.52}

This implies not only the absence of anomalies with ultraviolet dimensions
less than two, but also the absence of counterterms having infrared dimensions
less than two, thus avoiding from now on the problems deriving from the
infrared renormalization. Indeed, the general solution of equation~\equ{3.51}
is
\eq
{\cal A}^{(2)}={\cal D}_{\cal I}\widehat{\cal A}^{(2)}+
\widetilde{\cal A}^{(2)}\ .
\eqn{3.53}
A rapid investigation reveals that
\eq
r({\cal A}^{(2)})\geq 4
\eqn{3.54}
and, since the linearized Slavnov operator~${\cal D}_{\cal I}$ can raise
the infrared dimensions by at most two units, equation~\equ{3.54} implies
for the counterterm
\eq
r(\widehat{\cal A}^{(2)})\geq 2\ ,
\eqn{3.55}
as required by the infrared power counting. We are then left with the task of \
proving the absence of anomalies
\eq
\widetilde{\cal A}^{(2)}=0\ .
\eqn{3.56}

To study the cohomology problem in equation~\equ{3.51}, we filter the
operator~${\cal D}_{\cal I}$ with the filtering operator~\cite{dixon}
\eq
{\cal N} = \xi^\mu\pad{}{\xi^\mu}+\eta^\mu\pad{}{\eta^\mu}\ ,
\eqn{3.57}
according to which~${\cal D}_{\cal I}$ decomposes as
\eq
D_{{\cal I}}=D_{{\cal I}}^{(0)}+D^{(R)}\ ,
\eqn{deco}
where
\eq
D_{{\cal I}}^{(0)} = \widetilde{B}_{\Sigma_m} - \xi^\mu\pad{}{\eta^\mu}\ .
\eqn{3.59}
The nilpotency of the linearized massive Slavnov operator~\equ{2.9bis}
implies that
\eq
D_{{\cal I}}^{(0)} D_{{\cal I}}^{(0)} =0\ .
\eqn{3.60}

We know~\cite{dixon} that the
space of solutions~${\cal A}^{(2)}$ of the equation~\equ{3.51}
is isomorphic to a subspace of the solutions of the cohomology equation for
the operator~$D_{{\cal I}}^{(0)}$
\eq
D_{{\cal I}}^{(0)} X=0\ .
\eqn{3.61}

Now, the cohomology sector of the operator~~$D_{{\cal I}}^{(0)}$ coincides
with that of the linearized ordinary Slavnov operator
\eq
\widetilde{B}_{\Sigma_m}\Delta=0\ ,
\eqn{3.62}
where $\Delta$ does not depend on the global ghosts~$(\xi^\mu, \eta^\mu)$,
which appear in ~$D_{{\cal I}}^{(0)}$ as BRS doublets~\cite{dixon}.

Let us characterize $\Delta$ in terms of differential forms
\eq
\Delta=\int \Delta^1_2(x)\ .
\eqn{3.63}
Equation \equ{3.62} can be casted into a local one
\eq
\widetilde{B}_{\Sigma_m}\Delta^{1}_{2}(x) + d \Delta^{2}_{1}(x) =0\ ,
\eqn{3.64}
where $d$ is the exterior derivative and we adopted the usual notation
\eq
\Delta^{p}_{q}(x) \ \ ;\ \
\left\{
\begin{array}{lcl}
p &=& \mbox{\it ghost number}\\
q &=& \mbox{\it form degree}\\
\end{array}
\right.
\eqn{}

The descent equations~\cite{zumino,ps} deriving from~\equ{3.64} are
\eq
\widetilde{B}_{\Sigma_m} \Delta^{1}_{2} + d\Delta^{2}_{1}=0
\eqn{3.65}
\eq
\widetilde{B}_{\Sigma_m} \Delta^{2}_{1} + d\Delta^{3}_{0}=0
\eqn{3.66}
\eq
\widetilde{B}_{\Sigma_m} \Delta^{3}_{0}=0\ .
\eqn{3.67}

The general solution of equation~\equ{3.67} is
\eq
\Delta^{3}_{0}=\sum_{n=0}^{+\infty}a_{(n)}^{[abc]P_n}c^ac^bc^c\Phi^{P_n} +
\widetilde{B}_{\Sigma_m}\Delta^2_0\ ,
\eqn{3.68}
where $\left\{a_{(n)}^{[abc]P_n}\right\}$ is a sequence of invariant tensors.

The cocycle $\Delta^3_0$ yields as candidate for the anomaly of the
operator~${\cal D}_{\cal I}$
\eq\ba{rl}
{\cal A}=&\!\!\sum_{n=0}^{+\infty}\intx
a_{(n)}^{[abc](p_1\ldots p_n)}\LP-
6\rho^ac^bc^c \phi^{p_1}\ldots\phi^{p_n} -
6\varepsilon^{\mu\nu}A^a_\mu A^b_\nu c^c \phi^{p_1}\ldots\phi^{p_n}
\\&\!\!+
6nA^a_\mu c^bc^c\widehat\Omega^{p_1\mu} \phi^{p_2}\ldots\phi^{p_n} -
2nc^ac^bc^cL^{p_1} \phi^{p_2}\ldots\phi^{p_n} \\&\!\!+
n(n-1)\varepsilon_{\mu\nu}c^ac^bc^c
\widehat\Omega^{p_1\mu}\widehat\Omega^{p_2\nu} \phi^{p_3}\ldots\phi^{p_n} \RP\
,
\ea\eqn{3.69}
which does not satisfy the ghost equation~\equ{3.45bis}, so that it must be
\eq
a_{(n)}^{[abc]P_n}=0\Longrightarrow\Delta^3_0=
\widetilde{B}_{\Sigma_m} \Delta^2_0\ .
\eqn{3.70}

Equation \equ{3.66} becomes now a problem of local cohomology
\eq
\widetilde{B}_{\Sigma_m} (\Delta^2_1-d\Delta^2_0)=0\ .
\eqn{3.71}
To solve it, we filter the operator $\widetilde{B}_{\Sigma_m}$~\equ{2.9bis}
with
\eq\ba{rl}
\widetilde{\cal N}=
\intx\LP &\!\!
A^a_\mu\fud{}{A^a_\mu} + \phi^a\fud{}{\phi^a} + c^a\fud{}{c^a} +
\widehat\Omega^{a\mu}\fud{}{\widehat\Omega^{a\mu}} \\
&\!\!
+ \widehat{L}^a\fud{}{\widehat{L}^a}+
\rho^a\fud{}{\rho^a}+
\tau_3^\mu\fud{}{\tau_3^\mu}+\tau_4^\mu\fud{}{\tau_4^\mu}\RP\ ,
\ea\eqn{3.72}
according to which $\widetilde{B}_{\Sigma_m}$ decomposes as
\eq
\widetilde{B}_{\Sigma_m} =\widetilde{B}_{\Sigma_m}^{(0)}+
\widetilde{B}_{\Sigma_m}^{(R)}\ ,
\eqn{3.73}
where
\eq
\widetilde{B}_{\Sigma_m}^{(0)}=\widehat{B}^{(0)}-(\tau_1+m^2)\pad{}{\tau_2}\ ,
\eqn{3.73bis}
and
\eq
\widehat{B}^{(0)}=\intx\LP
-\partial_\mu c^a \fud{}{A^a_\mu}+
\varepsilon_{\mu\nu}\partial^\nu\phi^a\fud{}{\widehat\Omega^{a\mu}}-
\partial\widehat\Omega^a\fud{}{L^a}+
\varepsilon^{\mu\nu}\partial_\mu A^a_\nu\fud{}{\rho^a}+
\tau_3^\mu\fud{}{\tau_4^\mu}\RP\ .
\eqn{3.74}
The operator $\widetilde{B}_{\Sigma_m}^{(0)}$ is nilpotent
\eq
\widetilde{B}_{\Sigma_m}^{(0)}\widetilde{B}_{\Sigma_m}^{(0)}=0\ .
\eqn{3.75}

Now, it is easy to show~\cite{dixon,ps} that the local
cohomology of~$\widetilde{B}_{\Sigma_m}^{(0)}$ can depend only on the
undifferentiated fields $(c^a, \phi^a, \tau_1)$ thus solving the
equations~\equ{3.65},~\equ{3.66} one finds
\eq
\Delta^1_2(x)=\widetilde{B}_{\Sigma_m}^{(0)}\Delta^0_2(x) +
d\Delta^1_1(x) +\widehat\Delta^1_2(c^a,\phi^a,\tau_1)\ .
\eqn{3.76}
Coming back to the integrated level and remembering that, due to the ghost
condition~\equ{3.45bis}, the anomaly cannot depend
on the undifferentiated ghost~$c^a$,
equation~\equ{3.76} implies that
the cohomology of~$\widetilde{B}_{\Sigma_m}$ is empty.

This concludes the proof of the absence of anomalies of the
operator~${\cal D}_{\cal I}$. Indeed, since the cohomology
of~${\cal D}_{\cal I}$ is isomorphic~\cite{dixon} to a subspace of that
of~$\widetilde{B}_{\Sigma_m}$, we can affirm that
\eq
{\cal D}_{\cal I}{\cal A}^{(2)}=0\Longrightarrow
{\cal A}^{(2)}={\cal D}_{\cal I}\widehat{\cal A}^{(2)}\ ,
\eqn{3.77}
which was our purpose.

\section{Conclusions}

The analysis we have presented shows that the infrared regularized (massive)
two--dimensional BF model shares with its higher dimensionality partners the
same ``topological'' peculiarity of being finite; hence the introduction
of a mass parameter is, in this respect, harmless.
Furthermore the classical relation~\equ{shift} is easily shown to hold
unaltered
to all orders of perturbation theory by a straightforward application of the
Quantum Action Principle~\cite{qap}, thanks to the fact that, lacking
any counterterm, the classical and the effective action coincide.
This proves that the Callan-Symanzik equation is completely trivial in this
model. The mass parameter is implicitely identified by the integrated equation
of motion of the ghost~\equ{2.17} which implies
\eq
\left.
\frac{\delta^2\Gamma}
{\delta{\bar{c}}^a(0)\delta\tilde{c}^a(0)}\right|_{\psi=0}=-m^2\ \ ,\ \
\psi=\mbox{all fields}\ ,
\eqn{5.1}
as a normalization condition at zero momentum and the tilda denotes the
Fourier transform.
The question remains of the zero mass limit of this model .
There is an analogous situation in the two-dimensional nonlinear coset G/H
sigma model,which also needs a mass term to be free of infrared singularities
and where Elitzur conjectured~\cite{eli}, as was later proven~\cite{bbb},
that the correlation functions of G invariant local operators have a well
defined zero mass limit; these are identified as the local observables of the
theory. In the present case, where there are no local observables, the
conjecture may be transferred to the non-local ones (Wilson loops).
Finally  the classical equations of motion, {\it i.e.} the zero
curvature constraint for the gauge field and the vanishing of the covariant
derivative for the scalar, imply that $\phi^a$ lives on a sphere; this suggests
that the infrared problem may appear only in perturbation theory, whereas
it may not be present in the ``true'' model due to the compactness of the field
manifold.

\vspace{2cm}
\begin{center}
\bf Acknowledgements
\end{center}
We would like to thank O.~Piguet, P.~Provero, and S.P.~Sorella for
useful discussions. N.M. also thanks the {\it D\'epartement de Physique
Th\'eorique de l'Universit\'e de Gen\`eve }
for hospitality during part of this work.

\end{document}